\documentclass[a4paper]{jpconf}
\usepackage{graphicx}
\usepackage{amssymb,amsmath,bm,bbm}
\usepackage{epsf}
\usepackage{epsfig}
\usepackage{afterpage}
\usepackage{longtable}
\usepackage{cite}
\usepackage{latexsym, mathrsfs}
\usepackage{axodraw}
\usepackage{color}

\begin{document}
\title{Patterns of Flavour Violation in a Warped Extra Dimensional Model with Custodial Protection}

\author{Stefania Gori}

\address{Physik Department, Technische Universit\"at M\"unchen, D-85748 Garching, Germany}

\ead{sgori@ph.tum.de}

\begin{abstract}
We present a particular warped extra dimensional model, where the flavour diagonal and flavour non-diagonal $Z$ boson couplings to left-handed down quarks are protected by the custodial symmetry $P_{LR}$.
After a brief introduction of the model and of its main theoretical motivations, we present a complete study of rare $K$ and $B$ meson decays, including $K^+\to \pi^+\nu\bar\nu$, $K_L\to\pi^0\nu\bar\nu$, $B_{s,d}\to \mu^+\mu^-$ and
$B_{s,d}\to X_{s,d}\nu\bar\nu$. In particular we restrict the parameter space of the model to the subspace which fits all quark masses, CKM mixing parameters and all the measured $\Delta F=2$ observables, keeping the Kaluza-Klein scale in the reach of LHC ($\sim(2-3)$TeV). There we show that, in addition to the one loop contribution of the Standard Model (SM), the dominating new physics contribution to the rare decays of $K$ and $B_{s,d}$ mesons is the tree level exchange of the $Z$ boson of the SM governed by {\it right-handed} couplings to down-type quarks. In order to reduce the parameter dependence, we study correlations between various branching ratios of $B$ and $K$ mesons and between $\Delta F=1$ and $\Delta F=2$ observables. The patterns that we find allow to distinguish this new physics scenario from the SM and can offer an opportunity to future experiments to confirm or rule out the model.
\end{abstract}

\section{Introduction}
Models with a warped extra dimension (WED) \cite{Randall:1999ee}, in which all the Standard Model (SM) fields are allowed to propagate in the bulk, offer a natural solution, or at least alleviation, to some important puzzles of particle physics. As important examples we can mention the problem of the gauge hierarchy, the problem in generating the measured large hierarchies in masses and mixings of the fermions of the SM and in suppressing Flavour Changing Neutral Current (FCNC) interactions. Of first interest is the study of the flavour changing transitions in the context of these type of models. What we intend to present here are the results obtained in \cite{Blanke:2008yr} for the rare decays of $K$ and $B$ mesons taking into account all the constraints already found analysing the $\Delta F=2$ transitions \cite{Blanke:2008zb} (in particular the constraint from the well measured $\epsilon_K$ observable, in combination with a not too large fine tuning) and imposing a Kaluza-Klein scale in the reach of LHC ($(2-3)$TeV). In particular we want to show that, once we have fixed $ \epsilon_K$ in accordance with the experiments, the model predicts some testable patterns and correlations not only between various $\Delta F=1$ observables but also between $\Delta F=1$ and $\Delta F=2$ observables.

\section{Why Warped Extra Dimensions?}
There are many reasons for which one could be led to study warped extra dimensional models. To analyse all these motivations is clearly beyond the scope of this work. We will just present two of the most important: the problem of the gauge hierarchy and how to generate the ordinary fermion mass and mixing hierarchies in the context of WED.

\subsection{The Gauge Hierarchy Problem}
It is natural to assume that there is a theory beyond the Standard Model which completes the SM at higher energy scales ($\Lambda_{NP}$) and which reproduces at energy scales much lower than $\Lambda_{NP}$ the particle content of the SM. In this Beyond the Standard Model (BSM) theory we would have two scales: the electroweak symmetry breaking (EWSB) scale (the vacuum expectation value of the Higgs boson $v$) or equivalently the mass of the Higgs boson $M_H$ and the scale of new physics ($\Lambda_{NP}$)\footnote{If we want a theory which includes gravity, we have to take as high energy scale the Planck scale. In this case $v/\Lambda_{NP}\approx 10^{-16}$.}. This type of assumption leads to the gauge hierarchy problem: how to explain that we have this enormous separation between the two scales? The problem would be already grave at tree level, but it is even worse if we consider one loop corrections to the mass of the Higgs, since we have in general enormous quantum corrections to the mass of the Higgs from every particle which couples directly or indirectly to the Higgs field, pushing $M_H$ closer to the high energy scale.

The most popular way to address the problem is supersymmetry. Here we want to show briefly how we could solve it just adding an extra dimension to the four usual ones, and putting in this space a non flat metric.

The main idea of WED \cite{Randall:1999ee} is to consider a space (the bulk) with five dimensions. In the bulk, instead of putting the Minkowski metric, we put a solution of the 5 dimensional Einstein equations

\begin{equation}\label{metric}
 ds^2=e^{-2ky}\eta_{\mu\nu}dx^\mu dx^\nu -dy^2\,,
\end{equation}

\noindent where $\eta_{\mu\nu}=diag(1,-1,-1,-1)$ is the usual Minkowski metric and $k$ a free parameter, fixed only by the phenomenology.

The additional dimension is compactified on the $S^1/Z^2$ orbifold, generating two Lorentz invariant branes, one at $y=0$ (called Planck or UV brane), the other at $y=L$ (called SM or IR brane). Regarding the field content, in the simplest WED scenario the Higgs field of the SM is put on the IR brane, while gravity lives on the UV brane. So it is natural to assume that the typical energy scales on the two branes are the Planck mass ($M_{pl}$) on the UV brane and the electroweak scale ($v$) on the IR brane. 

The background metric that we have introduced in the bulk is responsible for solving or at least alleviating the gauge hierarchy problem \cite{Randall:1999ee}. Indeed, if we consider a fundamental energy scale of $\Lambda_{fund}$ on the UV brane, then we receive, going toward the IR brane, an exponential suppression, due to the warping factor $e^{-ky}$. In particular, if we compute the evolution of the fundamental energy scale until the IR brane, we find

\begin{equation}
 \Lambda_{eff}(L)=e^{-kL}\Lambda_{fund}\,.
\end{equation}

\noindent Since the Higgs boson lives on the IR brane, this effective energy scale can be seen naturally as the EWSB scale $v$. 

So, thanks to the exponential factor present in the metric, in order to generate a large hierarchy between $\Lambda_{fund}$ and $ \Lambda_{eff}(L)$, we do not need an extremely large exponent $kL$. In numbers, if we take $\Lambda_{fund}$ equal to $M_{pl}$, we just need $kL\approx 30$.

So the warped metric in the bulk allows us to see the EWSB scale just as the exponential suppression of the Planck scale with a quite natural value for the warping factor. In this way we alleviate the problem of the gauge hierarchy.

 \subsection{The Flavour Problem}
Warped extra dimensional models  can also provide naturally a solution to the following question: Why are the masses and mixings of fermions of the SM so hierarchical? To answer this question we need some more basics on WED. In these type of models gauge and matter fields can propagate into the fifth dimension. Every time that we solve the equations of motion for the fields, instead of finding a unique solution, we find an infinity of particles: the {\it Kaluza-Klein tower} (KK) \cite{Gherghetta:2000qt} of particles. If a tower has got a zero mode, that is a mode with vanishing mass, this particle is a particle of the SM. All the other more massive particles are KK excitations of the particles of the SM. 

In particular, the bulk profiles of left-handed and right-handed fermionic zero modes depend strongly on their so called bulk mass parameters $c$ 

\begin{equation}\label{zerofermions}
 f_L^{(0)}(y,c)=\sqrt{\frac{(1-2c)kL}{e^{(1-2c)kL}-1}}e^{-cky}\,,\hspace{0.5cm}f_R^{(0)}(y,c)=f_L^{(0)}(y,-c)\,,
\end{equation}

\noindent with respect to the warped metric.
We remark that in general the bulk masses $c$ are not universal for different fermion flavours. Using these shape functions $ f_{L,R}^{(0)}(y,c)$ and the Higgs SM doublet residing on the IR brane, we can write down the effective 4D Yukawa couplings

\begin{equation}
 Y_{ij}^{u,d}=\lambda_{ij}^{u,d}\frac{e^{kL}}{kL}f_L^{(0)}(L,c_Q^i)f_R^{(0)}(L,c_{u,d}^j)\,,
\end{equation}

\noindent where $\lambda^{u,d}$ are the fundamental 5D Yukawa coupling matrices. Due to the exponential dependence of $Y^{u,d}$ on the bulk mass parameters $c_{Q,u,d}$, the strong hierarchies of quark masses and mixings can be traced back to $O(1)$ bulk masses and anarchic 5D Yukawa couplings  $\lambda^{u,d}$. In numbers, if we choose completely anarchic 5D Yukawas, warping factor $kL\approx 30$ and bulk masses of quarks with different flavour differing by $50\%$, we find 4D Yukawas differing by two orders of magnitude. In conclusion, the hierarchies of 4D Yukawas, and so of quark masses and mixings, is explained naturally by a pure geometrical approach, leading to a partial solution of the flavour problem. What is still missing would be a theory able to explain the several values for the bulk masses.

\section{The Model}\label{model}

\subsection{The Field Content}\label{field content}

We consider an $SU(3)_c\times SU(2)_L\times SU(2)_R\times U(1)_X\times P_{LR}$ gauge theory on a slice of AdS$_5$ \cite{Randall:1999ee} with the metric given in (\ref{metric}). The fifth coordinate is restricted to the interval $0\le y\le L$, and the KK scale lowered to $2.5$TeV in the reach of LHC\footnote{See the original paper \cite{Blanke:2008zb} and the writeup \cite{Björn} for the details on how to combine this low high energy scale with the constraint coming from the observable $\epsilon_K$.}.

In the electroweak sector, we consider the gauge symmetry \cite{Agashe:2003zs,Csaki:2003zu,Agashe:2006at}
\begin{equation}\label{eq:bulk-group}
O(4)\times U(1)_X\sim SU(2)_L\times SU(2)_R\times P_{LR}\times U(1)_X\,,
\end{equation}
where $P_{LR}$ is the discrete symmetry interchanging the two $SU(2)$ groups. These two additional groups of symmetries, $SU(2)_R\times P_{LR}$, are put in order to have custodial protection of the T parameter \cite{Csaki:2003zu} and of the $Zb_L\bar{b}_L$ coupling \cite{Agashe:2006at}\footnote{In the following we will see the generalization of this protection to couplings that are not flavour diagonal.}\,\footnote{An analysis of the flavour and electroweak sector in a WED model without custodial protection can be found in \cite{Casagrande:2008hr}.}. The gauge group in (\ref{eq:bulk-group}) is broken to the SM one on the UV brane ($y=0$), i.\,e.
\begin{equation}
SU(2)_L\times SU(2)_R\times P_{LR}\times U(1)_X\to SU(2)_L\times U(1)_Y
\end{equation}

\noindent and on the IR brane to:
\begin{equation}
SU(2)_L\times SU(2)_R\times P_{LR}\times U(1)_X\to SU(2)_V\times U(1)_X\,.
\end{equation}

From the enlarged gauge group there arise three new neutral electroweak gauge bosons
\begin{equation}
Z_H\,,\qquad Z'\,, \qquad A^{(1)}
\end{equation}

\noindent in addition to the SM $Z$ boson and photon, where the first two are linear combinations of the gauge eigenstates $Z^{(1)}$ and $Z_X^{(1)}$ \cite{Albrecht:2009}. Neglecting small $SU(2)_R$ breaking effects on the UV brane and corrections due to electroweak symmetry breaking, one finds

\begin{equation}
M_{Z_H}=M_{Z'}=M_{A^{(1)}} \equiv M_{KK}\approx 2.5{\rm TeV}\,.
\end{equation}

Without entering too much into details for the fermion sector, we give the main features which will be useful in the following when we will study the rare decays of $K$ and $B$ mesons. 
In order to preserve the discrete $P_{LR}$ symmetry, we decide to embed all the three generations of left-handed SM down quarks into a symmetric representation of the discrete symmetry. In this way, also the off-diagonal couplings $Zd_L^i\bar{d}_L^j$ are protected \cite{Blanke:2008zb} by the same mechanism seen in \cite{Agashe:2006at} for the diagonal $Zb_L\bar{b}_L$ coupling.

\subsection{FCNC at Tree Level}\label{FCNC}
Due to the non universality of the shape functions of the zero modes of fermions (\ref{zerofermions}), we have a different localization of the fermions with different flavour in the bulk, depending on their bulk masses. In particular heavy fermions will reside close to the IR brane, instead light fermions are closer to the UV brane.

A different behaviour is shown by the gauge bosons. Solving their equation of motion, we find that the zero modes have a flat shape function, instead the KK modes are localized towards the IR brane.

Schematically we can write the interaction between SM fermions and a generic gauge boson as

\begin{equation}
\Delta_{L,R}\propto\int_0^L dy\,e^{ky} \left[f^{(0)}_{L,R}(y,c_\Psi^i)\right]^2 g(y)\,,
\end{equation}

\noindent where $g(y)$ is the shape function of the gauge boson in question.

As a consequence of the localization of the gauge KK modes
towards the IR brane, their couplings to zero mode fermions are not flavour universal, but depend strongly on the relevant bulk mass parameters $c_{Q,u,d}^i$. After rotation to the fermion mass eigenbasis, then, in general complex, flavour changing couplings of the heavy gauge bosons $Z_H$, $Z'$ and $A^{(1)}$ are induced. Additionally, due to the mixing of the SM $Z$ boson with the heavy KK modes $Z^{(1)}$ and $Z^{(1)}_X$, also the $Z$ boson couplings become flavour violating already at tree level. Due to these new sources of flavour and CP violation beyond the CKM matrix, the model is clearly a model beyond Minimal Flavour Violation (MFV) \cite{Hall:1990ac,D'Ambrosio:2002ex,Buras:2000dm}.

\section{Rare Decays: Theoretical Background}
From this section on we study the rare decays of $K$ and $B$ mesons. In particular we focus on the decays\footnote{For the complete analysis that includes also the rare decays $B\rightarrow K\bar\nu\nu$, $K_L\rightarrow\pi^0 l^+ l^-$, $K_L\rightarrow\mu^+\mu^-$ see the original paper \cite{Blanke:2008yr}.}

\begin{equation}\label{decayslist}
K^+\rightarrow\pi^+\bar\nu\nu;\hspace{0.2cm}K_L\rightarrow\pi^0\bar\nu\nu;\hspace{0.2cm}B\rightarrow X_{s,d}\bar\nu\nu;\hspace{0.2cm}B_s\rightarrow\mu^+\mu^-\,.
\end{equation}

To clarify in what our analysis consists of, we consider the first decay in (\ref{decayslist}). As it is well know, in the SM this process $K^+\rightarrow\pi^+\bar\nu\nu$ does not arise at tree level, but only at the one-loop level. 

\begin{figure}[ht]
\begin{center}
\begin{minipage}{18.5pc}
\includegraphics[width=18.5pc]{oneloopSM2.eps}
\caption{\label{oneloopdiagramsSM}\it Feynman diagrams responsible for the decay $K^+\rightarrow\pi^+\bar\nu\nu$ in the SM (from \cite{Buras:2004uu}).}
\end{minipage}\hspace{4pc}%
\begin{minipage}{14.5pc}
\begin{center}
\begin{picture}(100,100)(0,0)
\ArrowLine(0,10)(30,40)
\ArrowLine(30,40)(0,70)
\Photon(30,40)(100,40){2}{6}
\Vertex(30,40){1.3}
\Vertex(100,40){1.3}
\ArrowLine(100,40)(130,70)
\ArrowLine(130,10)(100,40)
\Text(65,21)[cb]{{\Black{$Z,Z^\prime,Z_H$}}}
\Text(20,60)[cb]{{\Black{$s$}}}
\Text(20,13)[cb]{{\Black{$d$}}}
\Text(110,60)[cb]{{\Black{$\nu$}}}
\Text(110,13)[cb]{{\Black{$\nu$}}}
\end{picture}
\end{center}
\caption{\label{treelevelWED} \it Tree level contributions of $Z$, $Z^\prime$ and $Z_H$ to the $s\to d\nu\bar\nu$ effective hamiltonian.}
\end{minipage} 
\end{center}
\end{figure}

If we consider the Feynman diagrams in Fig.~\ref{oneloopdiagramsSM} as effective vertices for the decay in question, we find the following top quark contribution to the effective hamiltonian\footnote{For simplicity we omitted the SM contribution of the charm quark, which is however relevant in the process $K^+\rightarrow\pi^+\bar\nu\nu$.}

\begin{equation}\label{HeffSM}
\mathcal{H}_{eff}^{SM}\propto V_{ts}^*V_{td}X_{SM}\left(\bar{s} \gamma_\mu\left(1-\gamma_5\right) d\right)\left(\bar{\nu} \gamma^\mu\left(1-\gamma_5\right) \nu\right)\,.
\end{equation}

At this point some comments are worthwhile:

\begin{itemize}
\item we have dependence just on an operator of the type $(V-A)\otimes (V-A)$;
\item the function $X_{SM}$ is a real and universal function in flavour space;
\item all the quark dependence is factorized in a prefactor that is the product of two elements of the CKM matrix ($V_{ts}^*V_{td}$).
\end{itemize}

In WED the pattern for this decay has some new aspects. As already seen in Sec.~\ref{FCNC}, we have FCNC transitions already at tree level mediated by the $Z$ boson of the SM and by the other heavy KK gauge bosons $Z'$ and $Z_H$. So we have the additional Feynman diagrams at tree level shown in Fig.~\ref{treelevelWED} to add to those of the SM.
Computing this new class of Feynman diagrams, we find for the effective hamiltonian two contributions: one is of the same type of that one already found in the SM (\ref{HeffSM}) ($V_{ts}^*V_{td}X^{V-A}\left(\bar{s} \gamma_\mu \left(1-\gamma_5\right)d\right)\left(\bar{\nu} \gamma^\mu\left(1-\gamma_5\right) \nu\right)$) with a function $X^{V-A}$ that is now complex and not universal anymore, and an additional one of the type

\begin{equation}
\mathcal{H}_{eff}^{new}\propto V_{ts}^*V_{td}X^V\left(\bar{s} \gamma_\mu d\right)\left(\bar{\nu} \gamma^\mu\left(1-\gamma_5\right) \nu\right)\,.
\end{equation}

\noindent Therefore,

\begin{itemize}
\item we have a new operator involved of the type $V\otimes (V-A)$;
\item the function $X^V$ is complex and not universal.
\end{itemize}

Interesting is also to notice that, computing all the three diagrams in Fig.~\ref{treelevelWED}, the main contribution is due to the coupling of the $Z$ boson of the SM to the {\it right-handed} down quarks. Instead the coupling of $Z$ to left-handed down quarks is strongly suppressed because of the protection already discussed in Sec.~\ref{model}.

\subsection{Bigger Contribution of New Physics in $K$ or $B$ Decays?}\label{sec:morenewphysics}
Summing the contribution of the SM and the new contributions of WED, we can write down in all generality the contribution of the top quark to the effective hamiltonian for the elementary process $q_1\rightarrow q_2\bar\nu\nu$

\begin{eqnarray}\nonumber
\mathcal{H}_{eff}^{tot}&\propto& V_{tq_1}^*V_{tq_2}\left(X_{SM}+{X_{q_1,q_2}^{V-A}}\right)\left(\bar{q_1} \gamma_\mu\left(1-\gamma_5\right) q_2\right)\left(\bar{\nu} \gamma^\mu\left(1-\gamma_5\right) \nu\right)+\\\label{Hefftot}
&+& V_{tq_1}^*V_{tq_2}{X_{q_1,q_2}^{V}}\left(\bar{q_1} \gamma_\mu q_2\right)\left(\bar{\nu} \gamma^\mu\left(1-\gamma_5\right) \nu\right)\,,
\end{eqnarray}

\noindent where we can express the new functions $X_{q_1,q_2}^{V-A}$ and $X_{q_1,q_2}^{V}$ as

\begin{equation}\label{nonuniversality}
X_{q_1,q_2}^{V-A,V}\propto\frac{1}{V_{tq_1}^*V_{tq_2}}F^{V-A,V}\left(\Delta_L^{\nu\nu},{\Delta_{L,R}^{q_1,q_2}}\right)\equiv \frac{1}{\lambda_t^{q_1,q_2}}F^{V-A,V}\left(\Delta_L^{\nu\nu},{\Delta_{L,R}^{q_1,q_2}}\right)\,.
\end{equation}

Since the functions $F$ depend on the quarks involved in the decay, the flavour of the quarks does not enter only through the dependence on the elements of the CKM matrix. This should be contrasted to the case of the SM where these decays are governed by a \emph{flavour-universal} loop function $X_{SM}$ and the only flavour dependence enters through the CKM factors.

Now we specialize this effective hamiltonian to the case of $s\rightarrow d\bar\nu\nu$ and $b\rightarrow (d,s)\bar\nu\nu$, for the decays of $K$ and $B$ mesons respectively. As $\lambda_t^{(s,d)}\simeq 4\cdot 10^{-4}$, whereas $\lambda_t^{(b,d)}\simeq 1\cdot 10^{-2}$ and $\lambda_t^{(b,s)}\simeq
4\cdot 10^{-2}$, we would roughly expect the deviation from the SM functions in the $K$ system to be by an order of magnitude larger than in the $B_d$ system, and even by a larger factor than in the $B_s$ system.

 This strong hierarchy in factors $1/\lambda_t^{q_1,q_2}$ is only partially compensated by the opposite hierarchy in the $F$ functions \cite{Blanke:2008yr}, so that we expect that contributions of new physics are bigger in the $K$ decays than in the $B$ decays, even if not by an order of magnitude. This expectation will be confirmed by our numerical analysis in the following section.

\section{Numerical Analysis}

\subsection{Breaking of Universality and New Sources of CP Violation}\label{Sec:breakdown}
In the model discussed here the universality in the rare decays is generally broken, as clearly seen in formulae (\ref{Hefftot}) and (\ref{nonuniversality}). 

Moreover, defining the total $X_i$ functions as

\begin{eqnarray}
X_K&=&X_{SM}+X_{sd}^{V-A}+X_{sd}^V=\left|X_K\right|e^{i\theta_X^K}\,,\\
X_s&=&X_{SM}+X_{bs}^{V-A}+X_{bs}^V=\left|X_s\right|e^{i\theta_X^s}\,,\\
X_d&=&X_{SM}+X_{bd}^{V-A}+X_{bd}^V=\left|X_d\right|e^{i\theta_X^d}\,,\\
\end{eqnarray}

\noindent we find that these functions become complex quantities and their phases turn out to exhibit a non-universal behaviour.

Numerically we find roughly
\begin{eqnarray}\label{numbersforX}
0.60 \le \frac{|X_K|}{X_{SM}} \le 1.30\,, \qquad 
0.95 \le \frac{|X_s|}{X_{SM}} \le 1.08\,, \qquad
0.90 \le \frac{|X_d|}{X_{SM}} \le 1.12\,, 
\end{eqnarray}

\noindent implying that the CP-conserving effects in the $K$ system can be much larger than in the  $B_d$ and $B_s$ systems (as expected already in the previous section), where new physics effects are found to be small.

\begin{figure}[ht]
\begin{center}
\begin{minipage}{16.5pc}
\includegraphics[width=16.5pc]{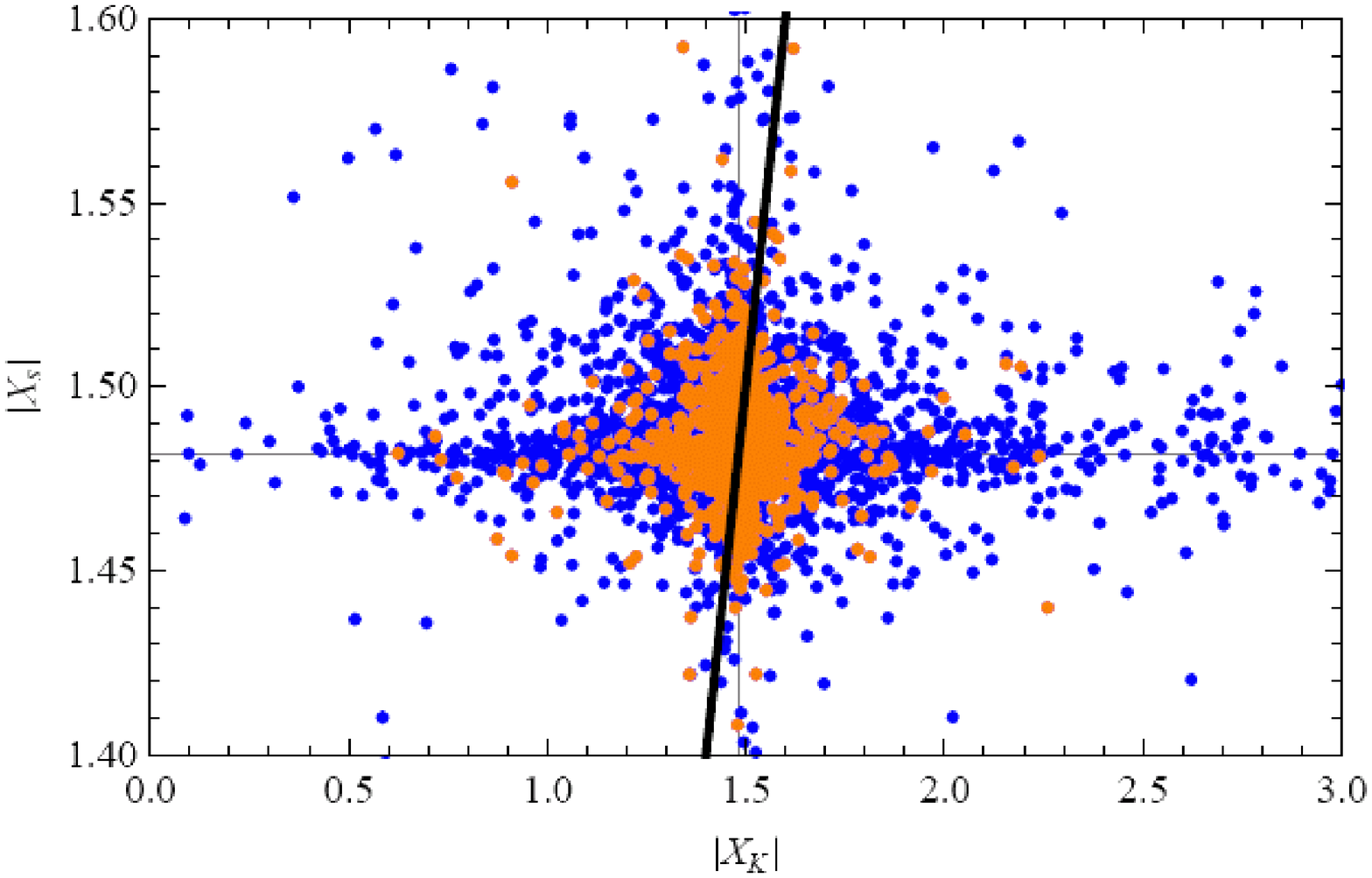}
\caption{\label{fig:breakdown}\it Breakdown of the universality between $|X_K|$ and $|X_s|$.
 Here and in the following we show in blue the points which satisfy all the constraints on the $\Delta F=2$ observables \cite{Blanke:2008zb} and in orange the points with in addition the requirement of moderate fine-tuning in $\varepsilon_K$ \\($\Delta_{BG}(\epsilon_K)\leq 20$ \cite{Barbieri:1987fn}).}
\end{minipage}\hspace{4pc}%
\begin{minipage}{16.5pc}
\includegraphics[width=16.5pc]{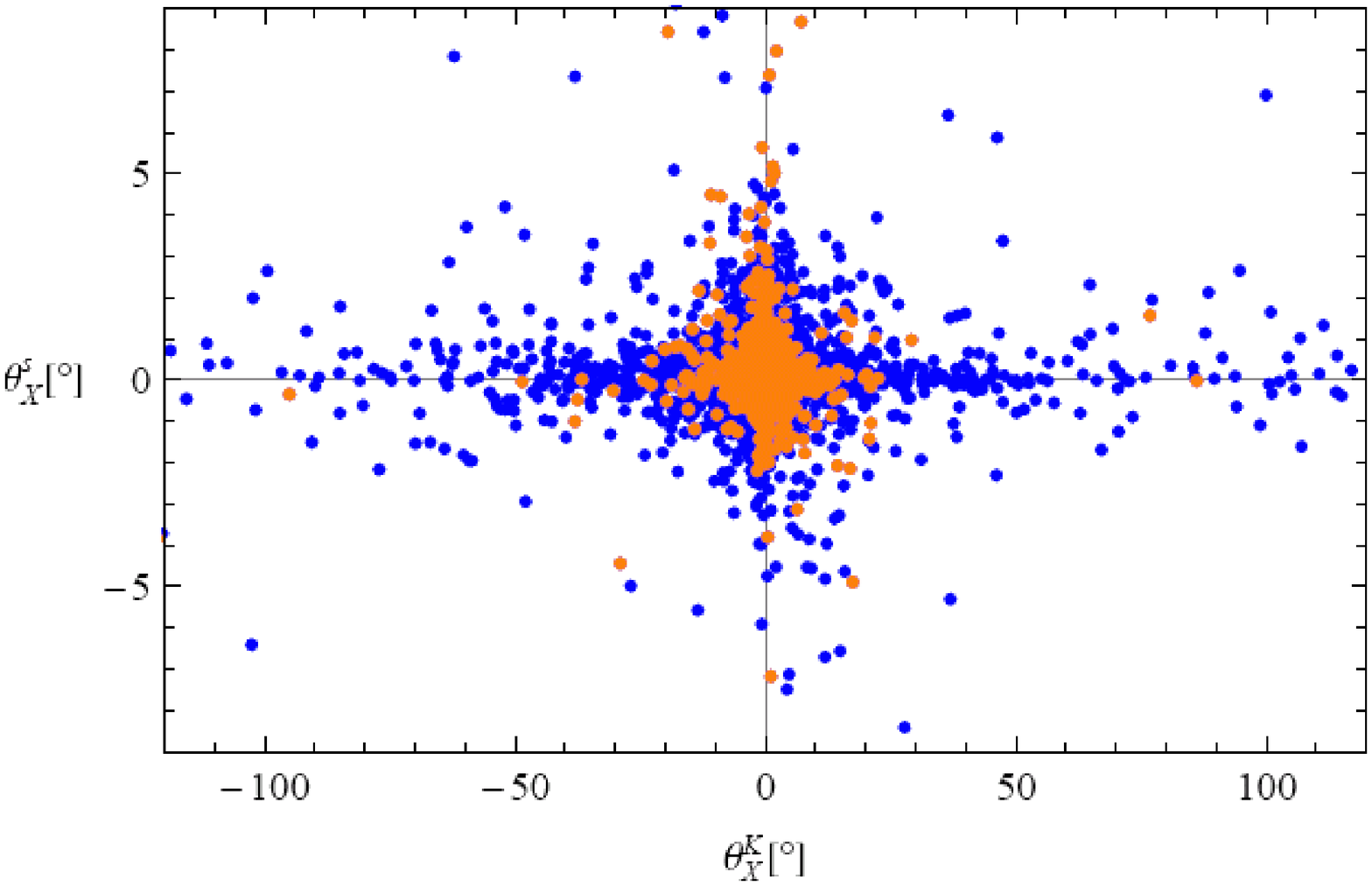}
\caption{\label{fig:newCPviolation}\it Breakdown of the universality between $\theta_X^K$ and $\theta_X^s$.}
\end{minipage} 
\end{center}
\end{figure}

We illustrate the first two quantities in (\ref{numbersforX}) in Fig.~\ref{fig:breakdown}, where we show the ranges allowed in the space $(|X_K|,|X_s|)$.
The solid thick line represents the relation $|X_s|=|X_K|$ that holds in the SM and more generally in models with Constrained Minimal Flavour Violation (CMFV) where all flavour violation is governed by the CKM matrix and only SM operators are relevant \cite{Buras:2000dm}). The crossing point of the thin solid lines indicate the SM value. The departure from the solid thick line gives the size of non-CMFV contributions that are caused dominantly by NP effects in the $K$ system. 

If we now investigate the new phases, i.e. new sources of CP violation, we find the ranges

\begin{equation}
-45^\circ \le \theta_X^K \le 25^\circ\,, \qquad 
-9^\circ \le \theta_X^d\le 8^\circ\,,\qquad
-2^\circ \le \theta_X^s\le 7^\circ\,,
\label{newCPviolation}
\end{equation}

\noindent implying that the new CP--violating effects in the $\Delta F=1$ $b\rightarrow d\bar\nu\nu$ and $b\rightarrow s\bar\nu\nu$ transitions are very small, while those in $K$ decays can be sizable (see also Fig.~\ref{fig:newCPviolation}).

Also from these last results it is evident that flavour universality can be significantly violated.

\subsection{Rare Decays of $K$ and $B$ Mesons}
In this subsection we investigate the possible correlations between different $\Delta F=1$ and also between $\Delta F=1$ and $\Delta F=2$ observables.
In the SM and in models with CMFV the rare decays analysed in the present work depend basically on universal functions. Consequently, a number of correlations exists between various observables not only within the $K$ and $B$ systems but also between $K$ and $B$ systems. In this subsection we analyse how some of these correlations are violated in the WED framework.

\begin{figure}[ht]
\begin{center}
\begin{minipage}{16.5pc}
\includegraphics[width=16.5pc]{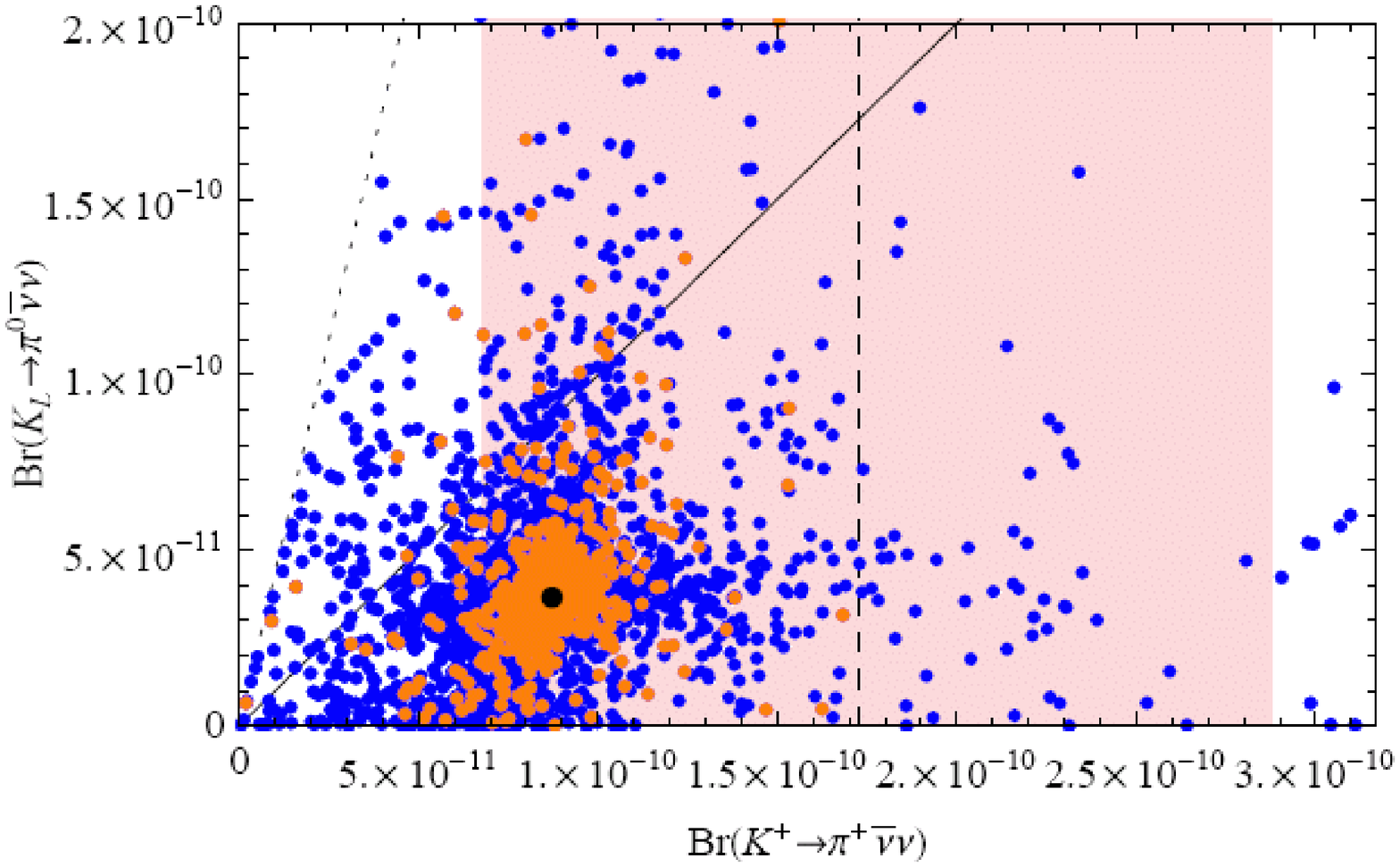}
\caption{\label{fig:KLKp}\it $Br(K_L\rightarrow\pi^0\bar\nu\nu)$ as a function of $Br(K^+\rightarrow\pi^+\bar\nu\nu)$. The shaded
    area represents the experimental $1\sigma$-range for $Br(K^+\rightarrow\pi^+\bar\nu\nu)$. The
    Grossman-Nir bound \cite{Grossman:1997sk} is displayed by the dotted line, while the solid line
    separates the two areas where $Br(K_L\rightarrow\pi^0\bar\nu\nu)$ is larger or smaller than
    $Br(K^+\rightarrow\pi^+\bar\nu\nu)$. The dark point shows the SM prediction.}
\end{minipage}\hspace{4pc}%
\begin{minipage}{16.5pc}
\includegraphics[width=16.5pc]{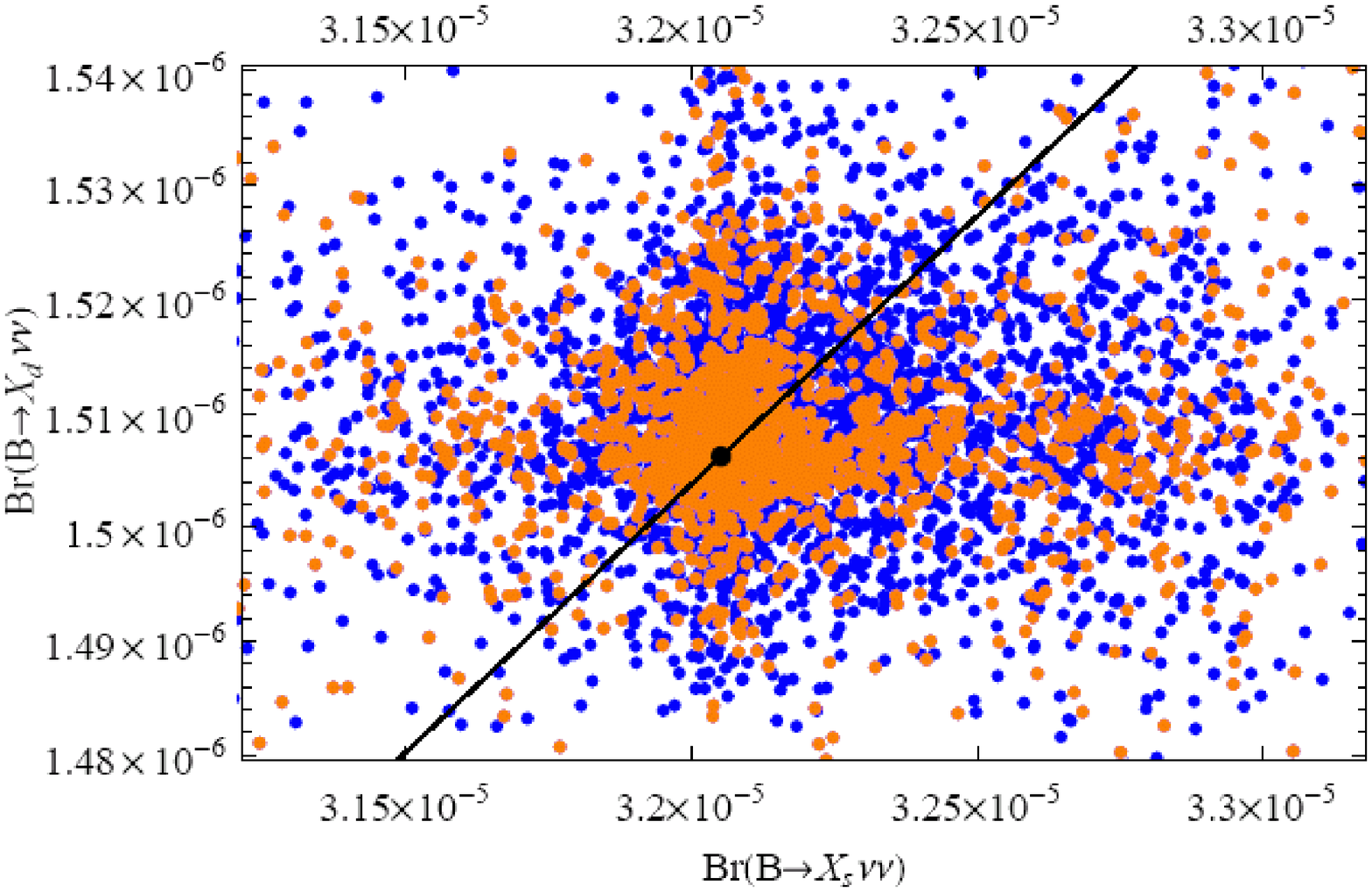}
\caption{\label{bsbd}\it Correlation between
$Br(B\to X_s\nu\bar\nu)$ and $Br(B\to X_d\nu\bar\nu)$. The black line represents the universal CMFV result given by the ratio $\frac{\left|V_{td}\right|^2}{\left|V_{ts}\right|^2}$ and the black point the SM prediction.}
\end{minipage} 
\end{center}
\end{figure}

First, we study the correlation between the branching ratios of the theoretical very clean decays $K_L\rightarrow\pi^0\bar\nu\nu$ and $K^+\rightarrow\pi^+\bar\nu\nu$. 
In Fig.~\ref{fig:KLKp} we show our result. We observe that both decays can differ simultaneously from the SM and in particular that the branching ratio $Br(K_L\rightarrow\pi^0\bar\nu\nu)$ can be as large as  $15\cdot 10^{-11}$, that is by a factor of 5 larger than its SM value \cite{Mescia:2007kn,Buras:2004uu}

\begin{equation}
Br(K_L\rightarrow\pi^0\bar\nu\nu)^{SM}=(2.8\pm 0.6)\cdot 10^{-11}\,,
\end{equation}

\noindent while being still consistent with the measured value for $Br(K^+\rightarrow\pi^+\bar\nu\nu)$. The latter branching ratio can be enhanced by at most a factor of 2 but this is sufficient to reach the central experimental value \cite{Artamonov:2008qb}

\begin{equation}
Br(K^+\rightarrow\pi^+\bar\nu\nu)^{exp}=(17.3^{+11.5}_{-10.5})\cdot 10^{-11}\,,
\end{equation}

\noindent to be compared with the SM value \cite{Brod:2008ss}

\begin{equation}
Br(K^+\rightarrow\pi^+\bar\nu\nu)^{SM}=(8.5\pm 0.7)\cdot 10^{-11}\,.
\end{equation}

Unfortunately, if we do the same type of analysis for the inclusive decays of $B$ mesons $B\rightarrow X_d \bar\nu\nu$ and $B\rightarrow X_s \bar\nu\nu$, we do not find the same spectacular deviation from the SM and in general from models with CMFV.

Of interest is the ratio

\begin{equation}\label{eq:P}
\frac{Br(B\rightarrow X_d\nu\bar\nu)}{Br(B\rightarrow X_s\nu\bar\nu)}=
\frac{\left|V_{td}\right|^2}{\left|V_{ts}\right|^2}\cdot P\,,
\end{equation}

\noindent where

\begin{equation}
P\equiv
\frac{|X_d^{V-A}+X_d^V/2|^2+|X_d^V/2|^2}
{|X_s^{V-A}+X_s^V/2|^2+|X_s^V/2|^2}\,.
\end{equation}

In the SM and models with CMFV we have a clear correlation between these branching ratios: $P=1$. Instead in WED we have deviations from this correlation, as seen in Fig.~\ref{bsbd}. In particular we find the range

\begin{equation}
0.93\le P \le 1.07\,.
\end{equation}

This result shows that NP effects in rare $B$ decays are significantly smaller than in rare $K$ decays as already expected from our anatomy of
NP effects in Sec.~\ref{sec:morenewphysics}. In $B$ decays the resulting deviation is small and will be difficult to measure.

As conclusion of this section we investigate one of the possible correlations between $K$ and $B$ decays (the two branching ratios of $K^+\rightarrow \pi^+\bar\nu\nu$ and $B_s\rightarrow\mu^+\mu^-$) and one between observables with $\Delta F=1$ and $\Delta F=2$ (the branching ratio of $K^+\rightarrow \pi^+\bar\nu\nu$ and the asymmetry $S_{\psi\phi}$).

\begin{figure}[ht]
\begin{center}
\begin{minipage}{16.5pc}
\includegraphics[width=16.5pc]{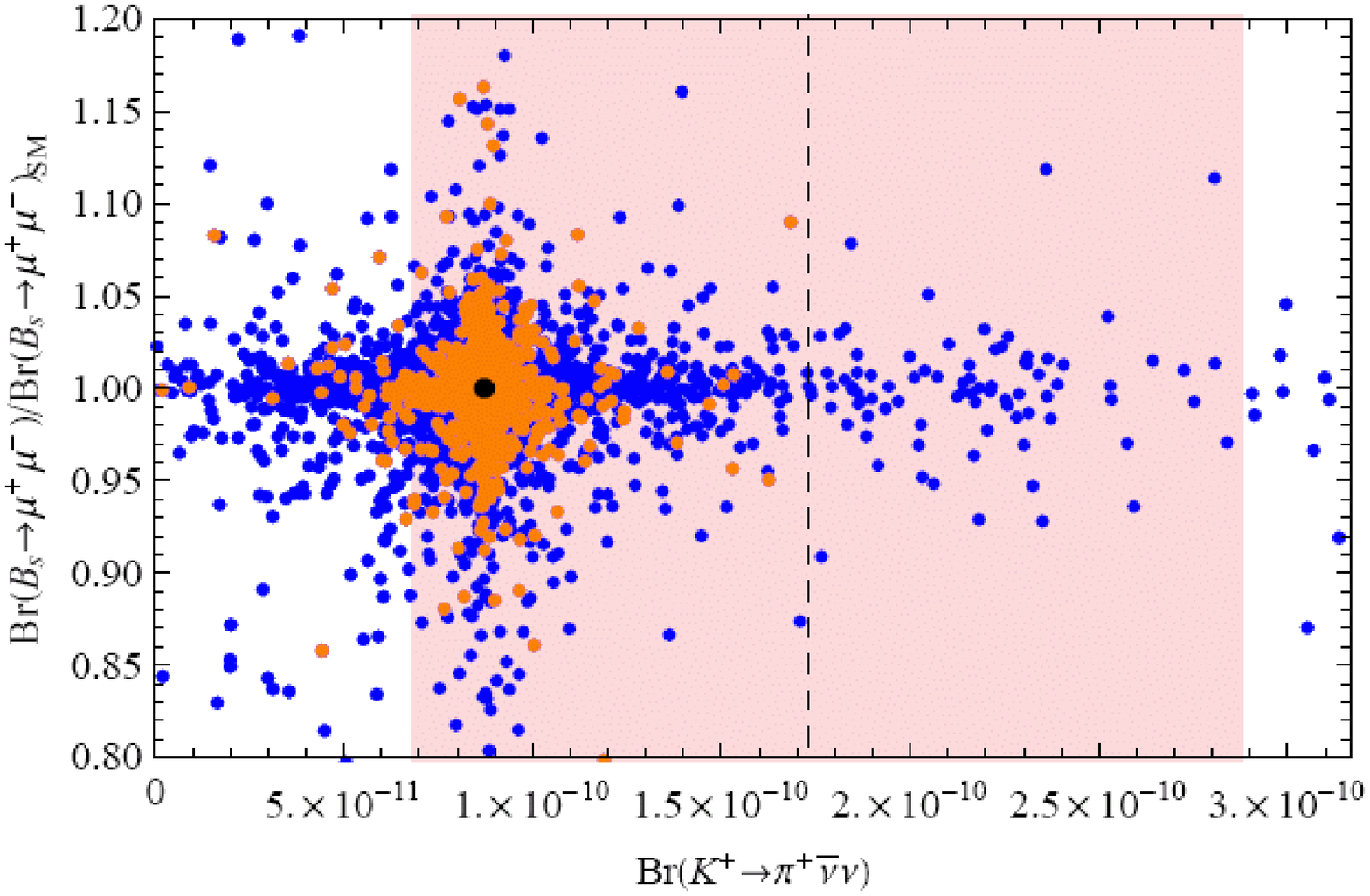}
\caption{\label{fig:BsKp}\it $Br(B_s\to \mu^+\mu^-)/Br(B_s\to \mu^+\mu^-)_\text{SM}$ as a
  function of $Br(K^+\rightarrow\pi^+\bar\nu\nu)$. The shaded area represents the experimental
  $1\sigma$-range for $Br(K^+\rightarrow\pi^+\bar\nu\nu)$ and the dark point shows the SM prediction.}
\end{minipage}\hspace{4pc}%
\begin{minipage}{16.5pc}
\includegraphics[width=16.5pc]{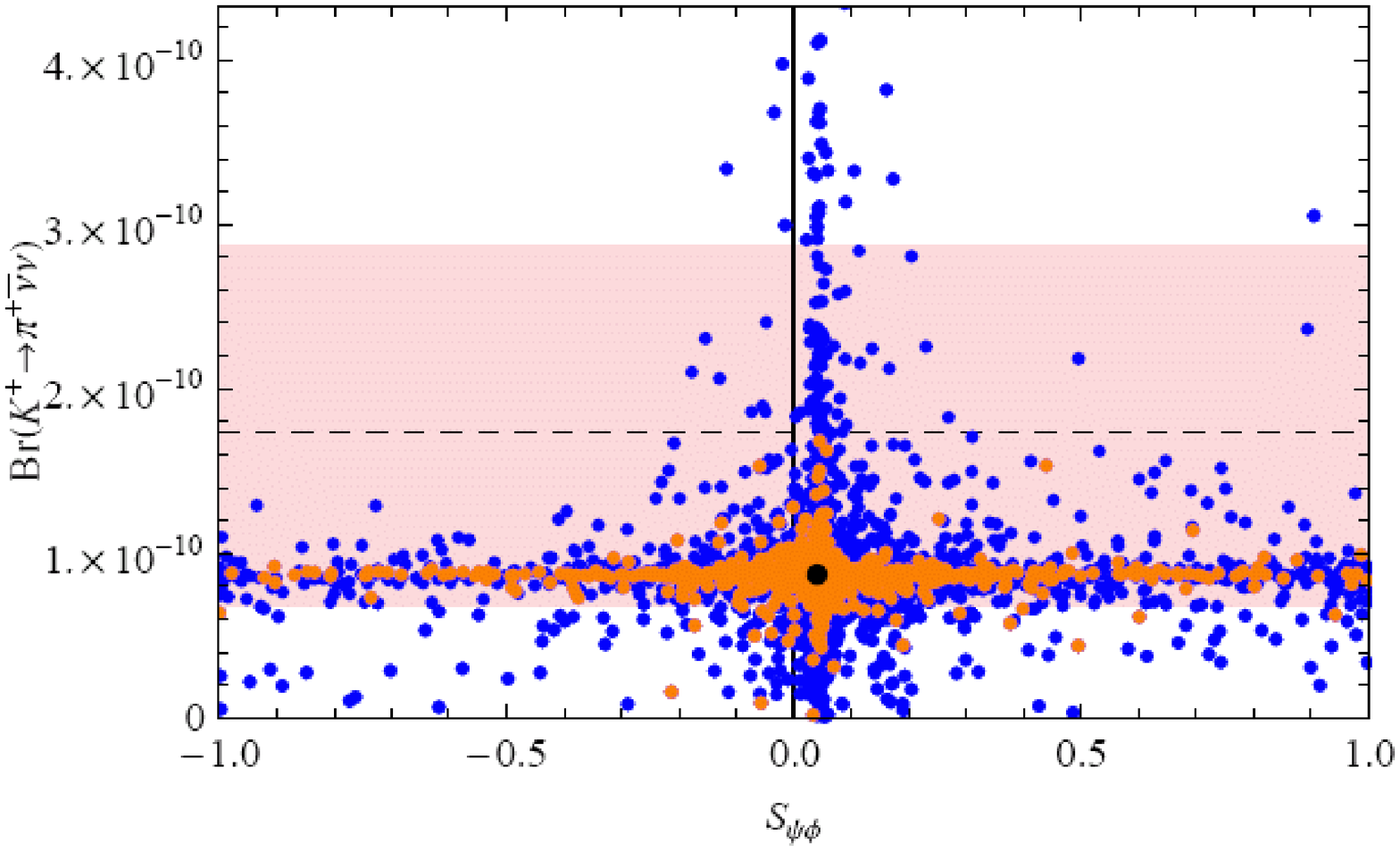}
\caption{\label{KpS}\it $Br(K^+ \rightarrow \pi^+
\nu \bar \nu)$ as a
  function of $S_{\psi \phi}$. The shaded area represents the experimental 
             $1\sigma$-range for 
             $Br(K^+ \rightarrow \pi^+
\nu \bar \nu)$. The dark point shows the SM prediction.}
\end{minipage} 
\end{center}
\end{figure}

In Fig.~\ref{fig:BsKp} we show that the branching ratio $Br(B_s\to \mu^+\mu^-)$ can deviate only by  $15\%$ from the SM value, while more pronounced effects ($200\%$) are possible in $Br(K^+\rightarrow\pi^+\bar\nu\nu)$ as we have seen before. When $Br(K^+\rightarrow\pi^+\bar\nu\nu)$ is sizably enhanced, it is almost impossible to distinguish the branching ratio $Br(B_s\to \mu^+\mu^-)$ from the SM prediction.

Instead in Fig.~\ref{KpS}, we observe that it is possible to obtain a large contribution of new physics for both the observables $S_{\psi\phi}$ and the branching ratio $Br(K^+ \rightarrow \pi^+\nu \bar \nu)$, but not simultaneously. 

In conclusion, Fig.~\ref{fig:BsKp} and Fig.~\ref{KpS} show that, despite of the many free parameters of the model, once that we have fixed all the $\Delta F=2$ observables and in particular $\epsilon_K$, the model predicts some precise patterns of flavour violation that can be confirmed or ruled out by future experiments. In particular, if future experiments will show a simultaneous enhancement for both $S_{\psi\phi}$ and $Br(K^+ \rightarrow \pi^+\nu \bar \nu)$ or big enhancements for rare $B$ decays, the model will be ruled out.

\section{Conclusions}
In the first part of this writing we have presented some of the reasons for which warped extra dimensional models could be a competitor of supersymmetry. In particular we showed that, in addition to addressing the gauge hierarchy problem, they can also naturally explain the hierarchies between masses and mixings of the quarks and leptons of the SM.

In the second part we have analysed the most interesting rare decays of $K$ and $B$ mesons in a warped extra dimensional model with a custodial protection of flavour diagonal and flavour non-diagonal $Z$ boson couplings to left-handed down quarks.

Once that we have fixed the scale of new physics to a scale that is in the reach of LHC ($(2-3)$TeV), we consider only the points of the parameter space which respect all the constraints from $\Delta F=2$ observables, and in particular from $\epsilon_K$. Studying this region of parameter space we arrive to these main conclusions:

{\it In the future experiments, an observation of a large $S_{\psi\phi}$ would, in the context of the model considered here, preclude sizable NP effects in rare $K$ decays. On the other hand, finding $S_{\psi\phi}$ to be SM-like will open the road to large NP effects in rare $K$ decays. Independently of the experimental value of $S_{\psi\phi}$, NP effects in rare $B$ decays are predicted to be small and an observation of large departures from SM predictions in future data would put the model considered here in serious difficulties.}

\section*{Acknowledgments}
I thank the other authors of \cite{Blanke:2008yr}: M. Blanke, A.J. Buras, B. Duling and K. Gemmler as well as
W. Altmannshofer and A.J. Buras for the careful reading of this writeup. This work was partially supported
by the European Community´s Marie Curie Research Training Network under contract MRTN-CT-2006-035505 (HEPTOOLS).

\section*{References}
\bibliographystyle{iopart-num}
\bibliography{proceeding1.bib}

\end{document}